\pdfoutput=1
\documentclass[a4paper, referee]{raa}

\usepackage{graphicx,times}             
\usepackage{natbib}
\usepackage{amssymb,amsmath}
\usepackage{upgreek}
\usepackage{lineno}
\bibpunct{(}{)}{;}{a}{}{,}

\usepackage[pagebackref=true]{hyperref}

\begin{document}

  \title{The Mini-SiTian Array: Optical design
}

   \volnopage{Vol.0 (20xx) No.0, 000--000}      
   \setcounter{page}{1}          

   \author{Zi-Jian Han
      \inst{1,2}
   \and Zheng-Yang Li
      \inst{1,2}
   \and Chao Chen
      \inst{1,2,3}
   \and Jia-Nan Cong
      \inst{1,2,3}
   \and Ting-Ting Liu
      \inst{1,2}
   \and Yi-Ming Zhang
      \inst{1,2,3}
   \and Qing-Shan Li
      \inst{4}
   \and Liang Chen
      \inst{1,2}
   \and Wei-Bin Kong
      \inst{1,2}
   }

   \institute{Nanjing Institute of Astronomical Optics \& Technology, Chinese Academy of Sciences, Nanjing 210042, China; {\it zyli@niaot.ac.cn}\\
        \and
             CAS Key Laboratory of Astronomical Optics \& Technology, Nanjing Institute of Astronomical Optics \& Technology, Nanjing 210042, China\\
        \and
             University of Chinese Academy of Sciences, Beijing 100049, China\\
        \and
             Astronomical Engineering Co. Ltd., Tianjin 300380, China \\
\vs\no
   {\small Received 20xx month day; accepted 20xx month day}}

\abstract{Time-domain astronomy is one of the most important areas. Large sky area, deep-field, and short timescale are the priority of time-domain observations. SiTian is an ambitious ground-based project processing all sky optical monitoring, aiming for sky-survey timescale of less than 1 day. It is developed by the Chinese Academy of Sciences, an integrated network of dozens of 1-m-class telescopes deployed worldwide. The Mini-SiTian Telescope Array is carried out for demonstrations on optical design, group scheduling, and software pipeline developments, to overcome the high technical and financial difficulties of SiTian project. One array contains three 300 mm F/3 telescope, with FOV of 5 degrees over 400-1000 nm wavelength range. The Mini-SiTian Telescope Array is now under commissioning in Xinglong Observatory, and a perfect platform for technical research and educational purposes.
\keywords{telescopes --- techniques: photometric --- methods: observational}
}

   \authorrunning{Z.-J. Han, Z.-Y. Li \& C. Chen et al.}            
   \titlerunning{The Mini-SiTian Array: Optical design }  

   \maketitle

%
%
\section{Introduction}           
\label{sect:intro}

Time-domain astronomy is one of the most important frontier areas. It aims to reveal the changes of various astronomical objects in the universe, as well as discover new astronomical objects and new phenomena through time-domain observations, including gravitational wave electromagnetic counterparts, exoplanet detection, and other popular directions in recent years. A series of influential large-scale time-domain survey projects such as The Panoramic Survey Telescope \& Rapid Response System (Pan-STARRS, \citealt{Kaiser+2002}), the Zwicky Transient Facility (ZTF, \citealt{Bellm+2019}), and the Legacy Survey of Space and Time (LSST, \citealt{Ivezic+2019}), are being carried out and planned to be implemented. Other ground-based sky-survey telescopes in China, such as the 2.5-m Wide-Field Survey Telescope (WFST, \citealt{LouZheng+2016}) and the Multi-Channel Photometric Survey Telescope (Mephisto, \citealt{YuanXiangYan+2020}) are also under commissioning in the past two years. But they are all facing the shortcomings of observing many astronomical objects and phenomena effectively with changing timescales within one day cadence. The cadence of LSST, ZTF, and WFST is about 2-4 days, and the cadence of Mephisto is about 2 weeks. Only one telescope will not satisfy all the needs of large sky area, deep field, and high cadance simultaneously for time-domain sky survey observations. High-precision measurements enable the precise extraction of useful information from photometric variability, especially in scenarios where the amplitude of variability is relatively weak (\citealt{LiShaSha+2022}). Consequently, high-precision photometric survey telescopes are of great significance to research in the field of time-domain astronomy, such as the study of supernovae, AGN, GRB, and TDEs (\citealt{FengHaiCheng+2021}, \citealt{SunTianRui+2024}). 
The SiTian project is a next-generation, large-scale time-domain survey designed to build an array of 60 optical telescopes, primarily located at observatory sites in China. This array will enable single-exposure observations of the entire northern sky with a cadence of only 30-minute, capturing true color ($gri$) time-series data down to about 21 mag. This project is proposed and led by the National Astronomical Observatories, Chinese Academy of Sciences (NAOC) \citealt{LiuJiFeng+2021}). The main science goals are the detection, identification, and monitoring of optical transients (such as gravitational wave events counterparts, fast radio bursts, and supernovae) on largely unknown timescales of less than 1 day. 
One SiTian node contains 3 prototype telescopes as an array (\citealt{ChenChao+2022}), which have high technical difficulty and high financial requirements in early development, so it is relatively difficult to carry out key software technology breakthroughs such as telescope control and group scheduling in a short time.
As the pathfinder for the SiTian project, the Mini-Sitian project utilizes an array of three 300 mm telescopes to simulate a single node of the full SiTian array. The Mini-Sitian has begun its survey since November 2022. One Mini-Sitian array is composed of three 300mm F/3 telescopes, with FOV of 5 degrees, and can observe in $g$, $r$, and $i$ band. The tracking accuracy is RMS $\le 0.5$ arcsecs in 10 minutes observations of stars, and RMS $\le 2.0$ arcsecs in observations of space targets.

In this article, the Mini-SiTian Telescope will be introduced. Instrument design will be described in Section ~\ref{sect:instrument}, instrument performances in laboratory test and commission will be shown in Section ~\ref{sec:Performance}, and the project will be summarized in Section 4.


\section{Instrument}
\label{sect:instrument}

Mini-SiTian Telescope is specifically designed for sky survey observations. With a large FOV and a fast focal ratio of F/3, the telescope uses a Cassegrain focus, where the detector is easier to mount on. The telescope has a good image quality on full FOV, which is capable of astrophotography, astronomical sky-survey, and space debris observations.

\subsection{Optical Design}

The optical system of the Mini-Sitian Telescope is heritage from the optical system of CSTAR (Chinese Small Telescope Array, \citealt{LiuGenRong+2009}). Evolved from the catadioptric optical system, it is a large field-of-view optical system that corrects the spherical aberration of the primary mirror at the entrance pupil to obtain a uniform star PSF distribution in the field of view. The telescope tube is compact with loose optomechanical tolerances. The disadvantage is that it is not easy to set up a straylight shield, leading to poor control of stray light, and thus requiring the addition of a front straylight shield. Key parameters of the telescope are as Table ~\ref{table:parameter}.

\begin{table}
\begin{center}
\caption[]{Key parameters of the Mini-Sitian Telescope.}\label{table:parameter}


 \begin{tabular}{clcl}
  \hline\noalign{\smallskip}
Parameter &  Value                   \\
  \hline\noalign{\smallskip}

Aperture                     & 300 mm      \\
Focal length               & 900 mm      \\
Focal ratio                & F/3          \\
Field of view              & 5 degrees   \\
Diameter of telescope tube & 420 mm      \\
Designed wavelength range  & 400-1000 nm \\
Weight  & 98 kg    \\
  \noalign{\smallskip}\hline
\end{tabular}
\end{center}
\end{table}

\begin{figure}
   \centering
   \includegraphics[width=12cm, angle=0]{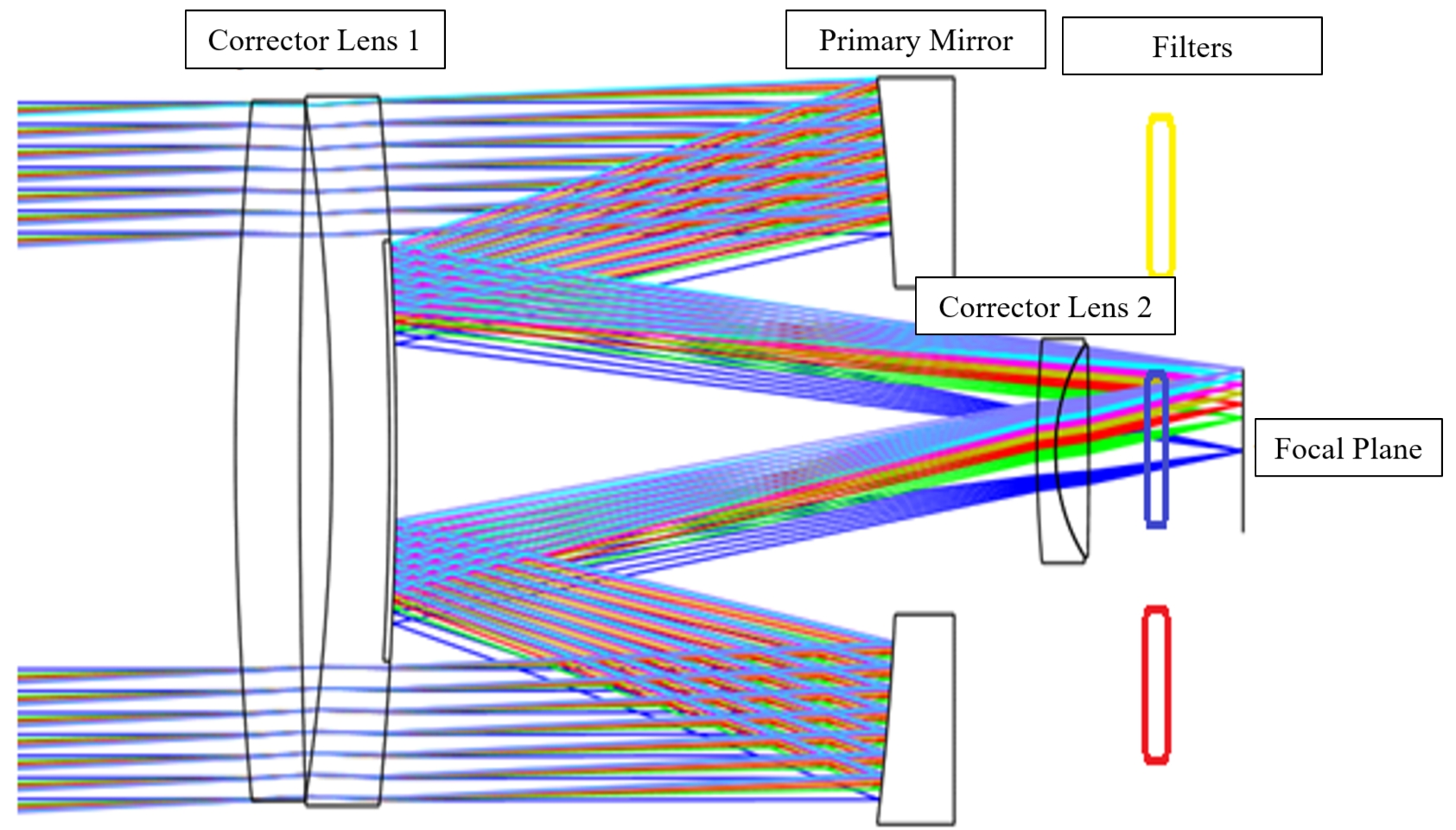}
   \caption{Optical layout of the Mini-SiTian Telescope.}
   \label{Fig1:layout}
   \end{figure}

The optical layout is shown in Figure ~\ref{Fig1:layout}. The corrector lens belonging to group 1 serves as the incoming block window of the telescope. Reflective coatings are made on the last surface to shorten the telescope tube. Corrector lens belonging to group 2 are used to enhance the image quality on marginal FOV. Three optical filters can be switched using a filter wheel. The size of focal plane is $\Phi$ 80 mm, matching a FOV of 5 degrees in diameter. 

\begin{figure}
   \centering
   \includegraphics[width=12cm, angle=0]{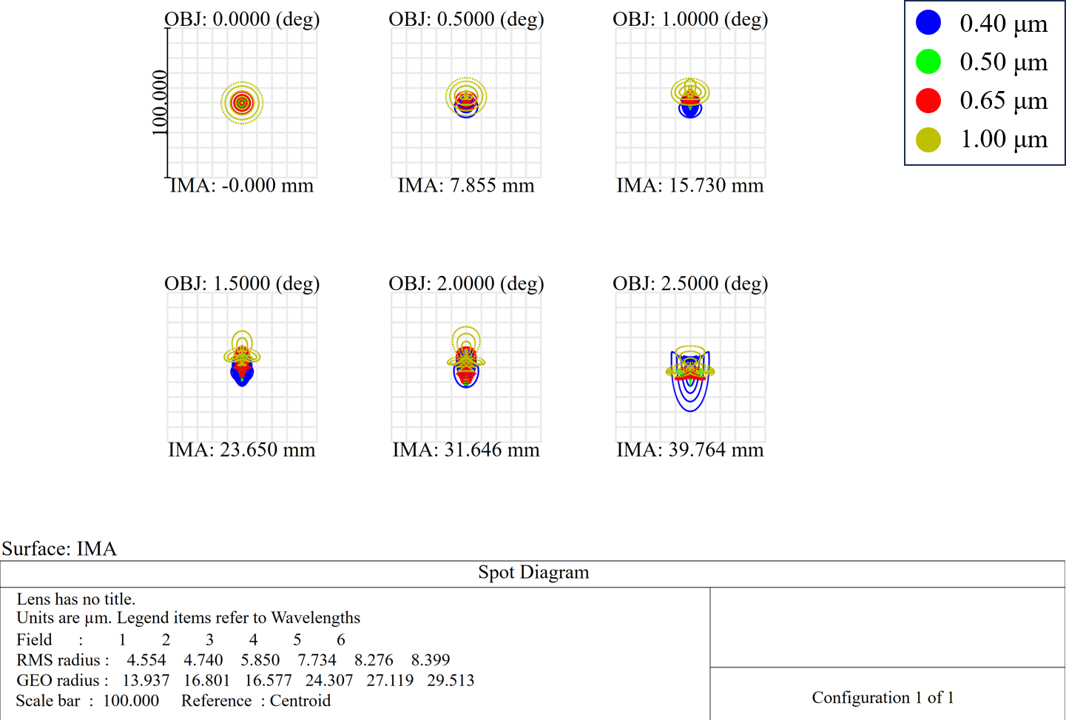}
   \caption{Spot diagram of the Mini-SiTian Telescope.}
   \label{Fig2:spotdiagram}
   \end{figure}

\begin{figure}
   \centering
   \includegraphics[width=12cm, angle=0]{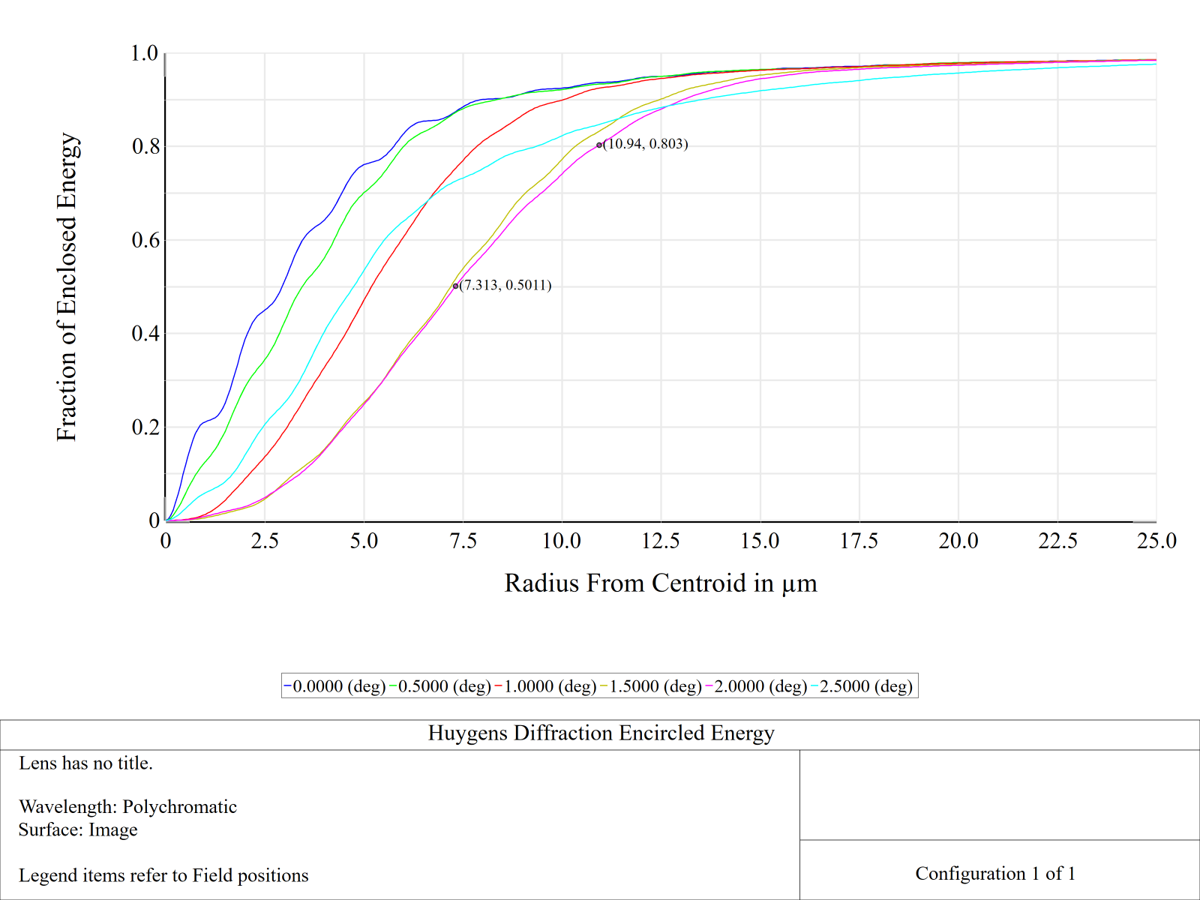}
   \caption{Diffraction encircled energy of the Mini-SiTian Telescope.}
   \label{Fig3:diffractionEE}
   \end{figure}

The optical spot diagram and the diffraction encircled energy diagram are shown in Figures ~\ref{Fig2:spotdiagram} and ~\ref{Fig3:diffractionEE}. The root-mean-square (RMS) radius is 8.728 $\upmu$m at the maximum FOV of 5 degrees over 400-1000 nm wavelength range. The diameter of 80\% encircled energy (EE80) is less than 5.04 arcsecs ($\sim$ 10.94 $\upmu$m). FWHM is less than 2.89 arcsecs ($\sim$ 6.29 $\upmu$m). 

The telescope employs focusing mechanisms to compensate for defocus caused by transportation, temperature shifts, and optical path differences while switching waveband filters. Different waveband filters, $g$, $r$, and $i$ can be inserted into the optical path for multi-band observations. 

\begin{figure}
   \centering
   \includegraphics[width=12cm, angle=0]{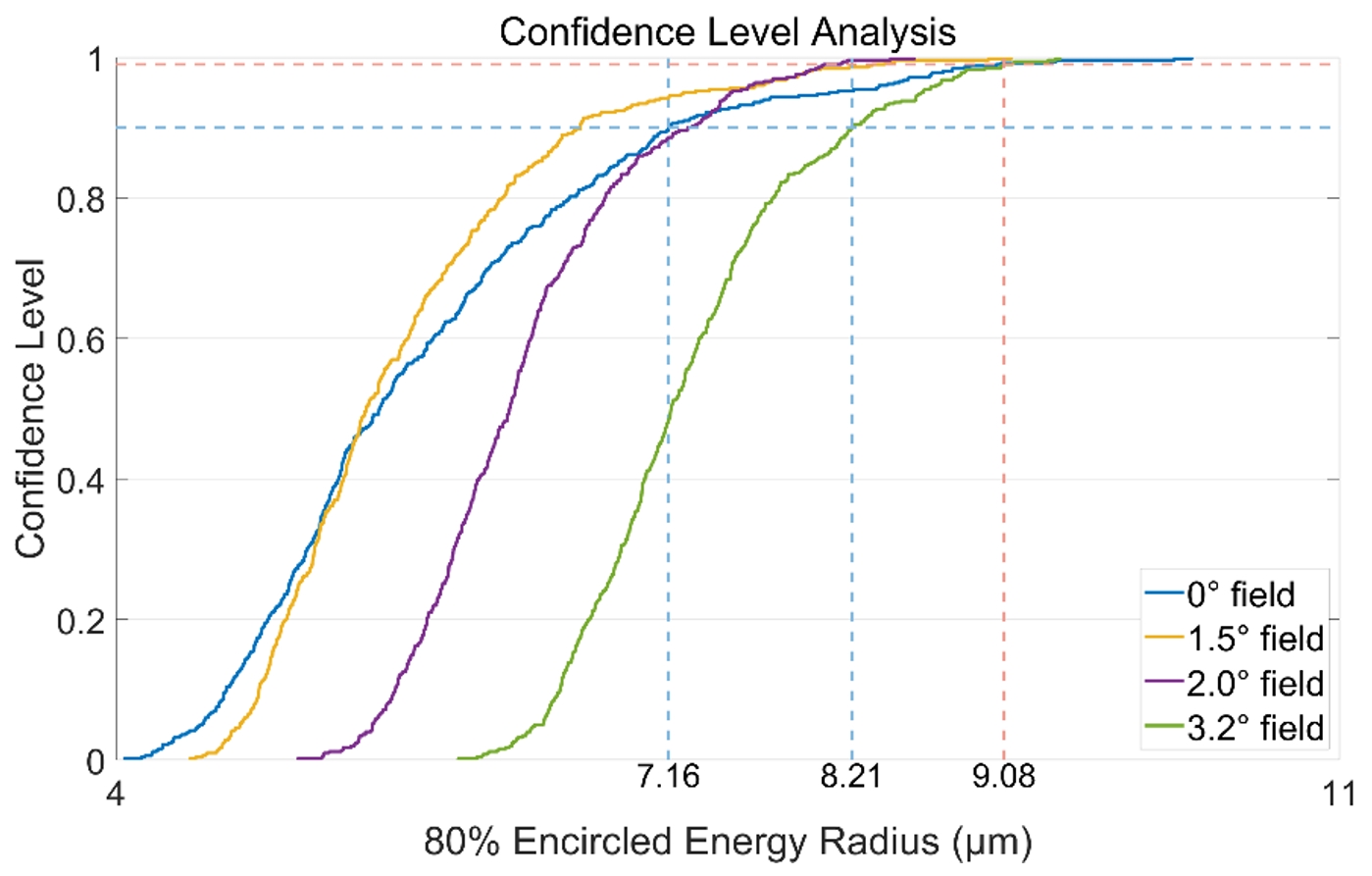}
   \caption{Results of the error budget analysis.}
   \label{Fig4:errorBudget}
   \end{figure}

Error budget analysis is essential in optical design to evaluate tolerances in manufacturing and mounting. Monte-Carlo analysis is performed on different FOVs to analysis the EE80 under different tolerances with ZEMAX software, which is shown in Figure ~\ref{Fig4:errorBudget}. The results indicated that, at the central field with a 90\% probability, the EE80 diameter is approximately 7.16 $\upmu$m (3.31 arcsecs). Furthermore, the EE80 diameter of the MST, under a 90\% probability, was smaller than 8.21 $\upmu$m (3.79 arcsecs) for the maximum FOV, and the EE80 diameter of MST, under a 99\% probability, was approximately 9.08 $\upmu$m (4.19 arcsecs) for the maximum FOV. These two values are extremely close, and this analysis confirmed that both the manufacturing process and the alignment are feasible. 

As for the primary and secondary mirrors, enhanced aluminum coating is implemented. The average reflectance of the enhanced aluminum coating is 89.6\% over 400 – 1000 nm. Anti-reflectance coatings are coated on the corrector lens, with an average transmittance of 97.4\% over 400 – 1000 nm. By calculating all the working surfaces of the telescope optics, the overall efficiency of the telescope is about 68.5\% over 400 – 1000 nm range, with minimum and maximum values of 64\% and 74\%, respectively.

\subsection{Stray Light Analysis}
Stray light refers to the light rays in an optical system that diffuse to the detector or imaging surface other than the target light rays and imaging light rays, as well as the target light rays that reach the detector through abnormal optical paths. Stray light degrades the contrast and modulation transfer function (MTF) of the image plane, resulting in reduced image layers, deteriorated clarity, and disordered energy distribution. Furthermore, stray light can generate light spots on the image plane, or even completely overwhelm the target signal with stray radiation noise. Therefore, it is necessary to suppress stray light of the Mini-SiTian Telescope.

The suppression level of stray light in optical-mechanical systems is generally evaluated using Point Source Transmittance (PST). PST is defined as the ratio of the irradiance reaching the detector surface after the radiation from a point source target at an off-axis angle of $\theta$ at wavelength $\lambda$ outside the system's field of view passes through the optical-mechanical system $E_{\mathrm{image}}(\theta, \lambda)$, to the irradiance of that light source at the system's entrance aperture $E_{\mathrm{ea}}(\theta, \lambda)$, which is 

\begin{equation}\label{eq:PST}
  \mathrm{PST}=\frac{E_{\mathrm{image}}(\theta, \lambda)}{E_{\mathrm{ea}}(\theta, \lambda)}
\end{equation}

\begin{figure}
   \centering
   \includegraphics[width=12cm, angle=0]{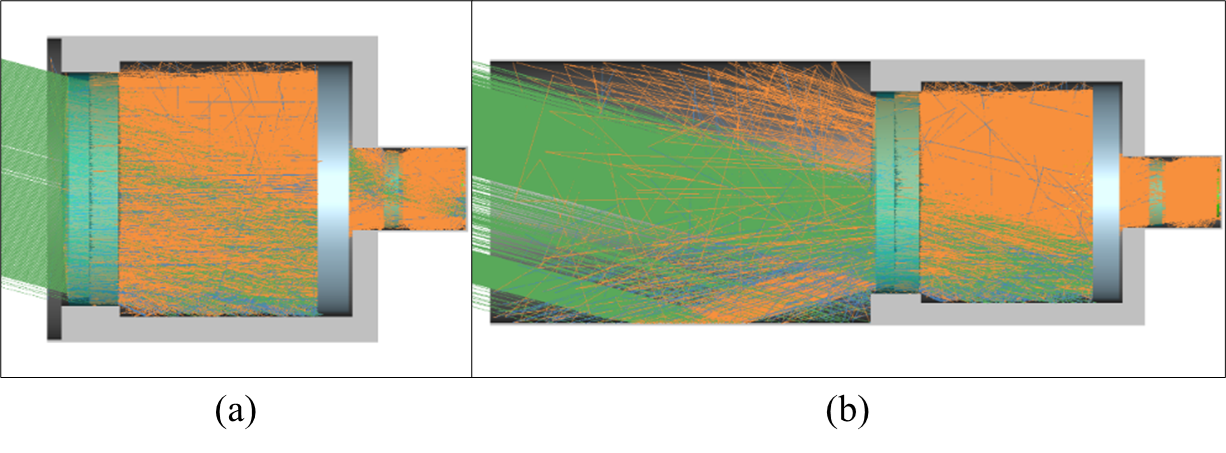}
   \caption{Stray light suppression (a) with and (b) without a front baffle of 60cm in length.}
   \label{Fig5:StrayLight}
   \end{figure}

Due to the compact structure and large field of view of the telescope, it is difficult to place a large baffle inside the telescope to suppress stray light. Especially when the incident angle of stray light is small, the telescope image is severely affected by stray light, leading to obvious ghost images. To more effectively suppress stray light, a long baffle needs to be installed in front of the telescope. Figures ~\ref{Fig5:StrayLight}(a) and ~\ref{Fig5:StrayLight}(b) show the stray light analysis optical-mechanical models without and with the front baffle, respectively. 

\begin{figure}
   \centering
   \includegraphics[width=12cm, angle=0]{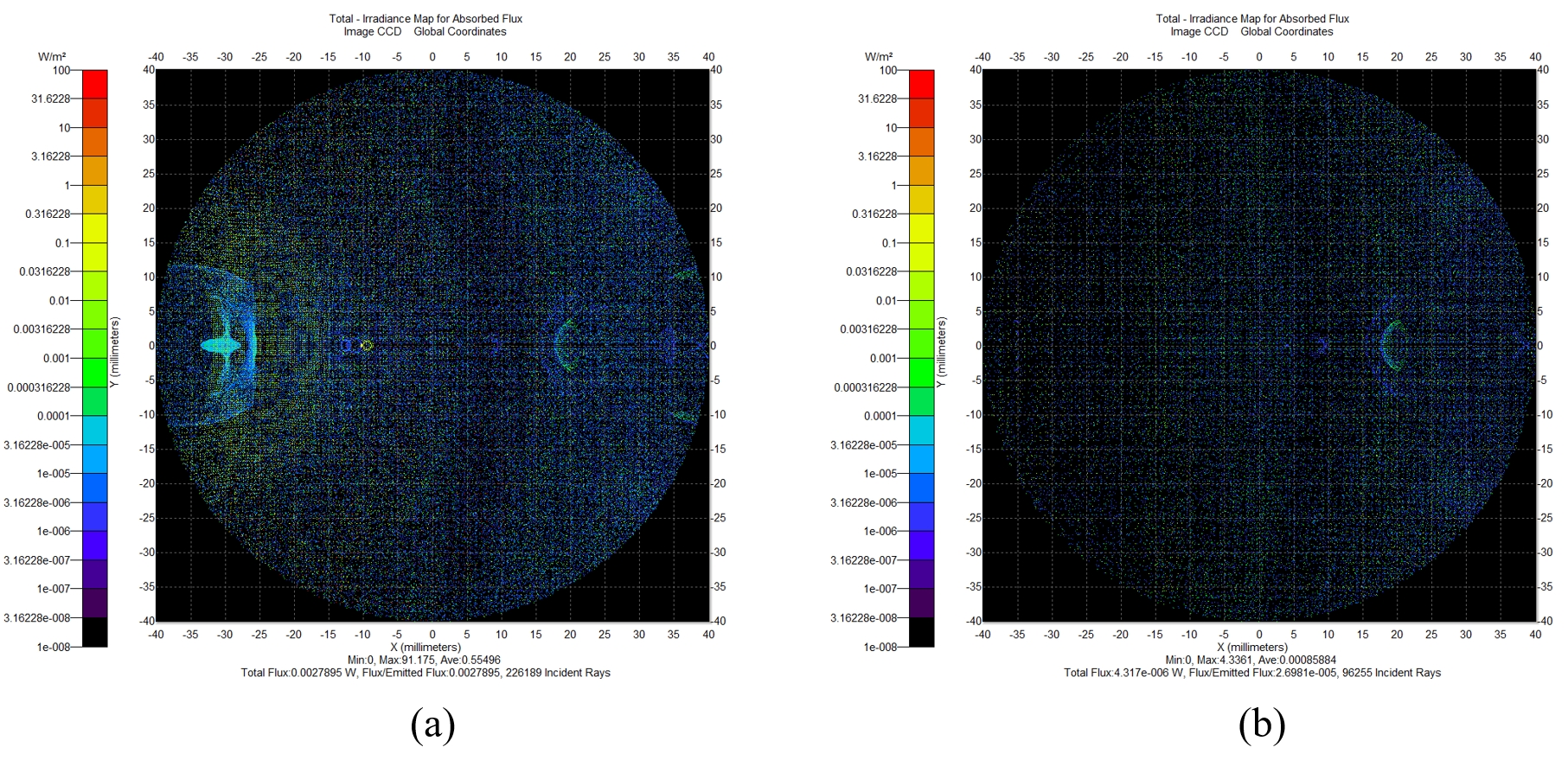}
   \caption{Distribution of stray light on the image plane (a)without a front baffle and (b) with a 60cm front baffle, both under an irradiance of $1\mathrm{W}/\mathrm{m}^{2}$ for the stray light source.}
   \label{Fig6:StrayLightDistribution}
   \end{figure}

Figures ~\ref{Fig6:StrayLightDistribution}(a) and ~\ref{Fig6:StrayLightDistribution}(b) present the illuminance of stray light on the image plane at a 15$^{\circ}$ incident angle without and with the front baffle, respectively, where the irradiance of the stray light source is $1\mathrm{W}/\mathrm{m}^{2}$. When no front baffle is used, the stray light brightness on the image plane is high, and a bright ghost image appears on the left side of the image plane. By using the front baffle, the stray light brightness and ghost image on the image plane are effectively suppressed. 

\begin{figure}
   \centering
   \includegraphics[width=12cm, angle=0]{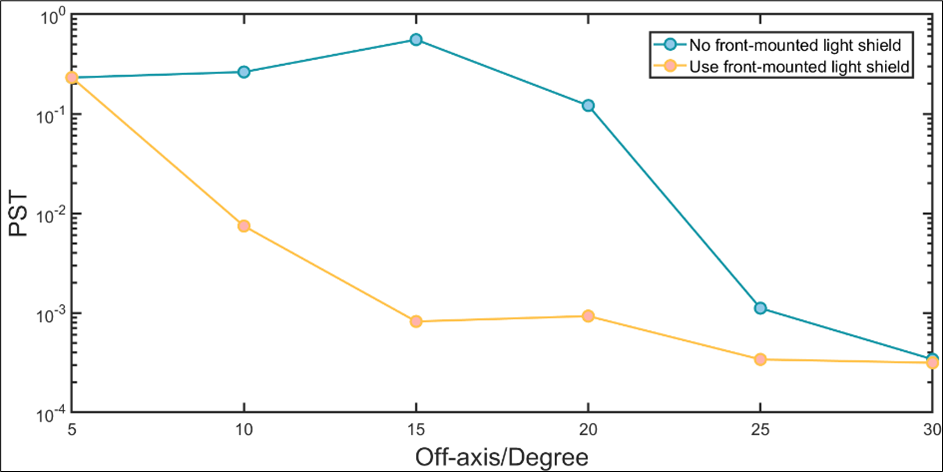}
   \caption{PST curves for the cases with and without a front baffle.}
   \label{Fig7:PSTCurve}
   \end{figure}

Figure ~\ref{Fig7:PSTCurve} shows the PST curves for both cases. Without the baffle, the incident angle of stray light needs to be above 30$^{\circ}$ for the system's PST to decrease to the order of $10^{-4}$; with the front baffle, this requirement is reduced to 15$^{\circ}$.

\subsection{Equatorial Mount}

\begin{figure}
   \centering
   \includegraphics[width=12cm, angle=0]{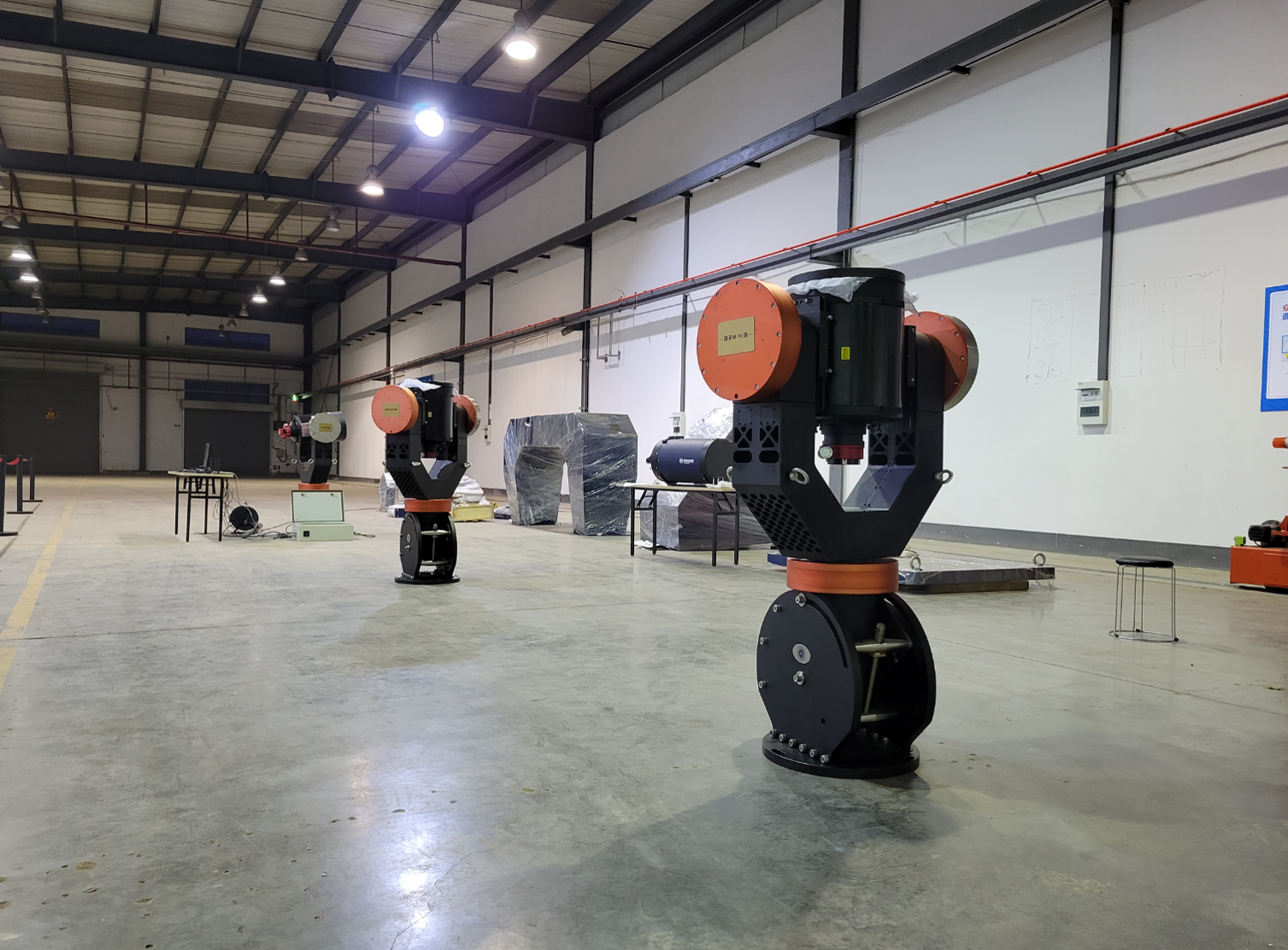}
   \caption{Equatorial mount of the Mini-SiTian Telescope Array in the factory.}
   \label{Fig8:EquatorialMount}
   \end{figure}

The equatorial mount is specially designed for sky-survey observations, as shown in Figure ~\ref{Fig8:EquatorialMount}. It uses Computerized Numerical Control (CNC) integrated processing and forming in manufacturing. To guarantee pointing and tracking accuracy, the equatorial mount uses an axial magnetic field torque motor and a high-precision industrial-grade 32-bit absolute encoder. The pointing accuracy is better than 10 arcsecs after pointing modelling, and the star-tracking accuracy is better than 0.5 arcsecs over 10 minutes. The equatorial mount can also be used for space targets observations, with tracking accuracy better than 2 arcsecs. 

The equatorial mount can be operated on either Windows or Linux platform, and it can also be operated on SkyView graphical users interface (GUI). It supports ASCOM RTS2 protocols, and third-party software such as SkyX. SDK and low-level commands are provided for further development purposes. 

As shown in Figure ~\ref{Fig6:StrayLightDistribution}, the equatorial mount can be quickly switched to alt-azimuth or equatorial configuration, without any lifting equipment. The telescope mount has a maximum loading capacity of 200 kg. The maximum power of the mount is 800W with a 200~240 VAC power supply.

\section{Performance}
\label{sec:Performance}

\subsection{Testing Observations}

\begin{figure}
   \centering
   \includegraphics[width=12cm, angle=0]{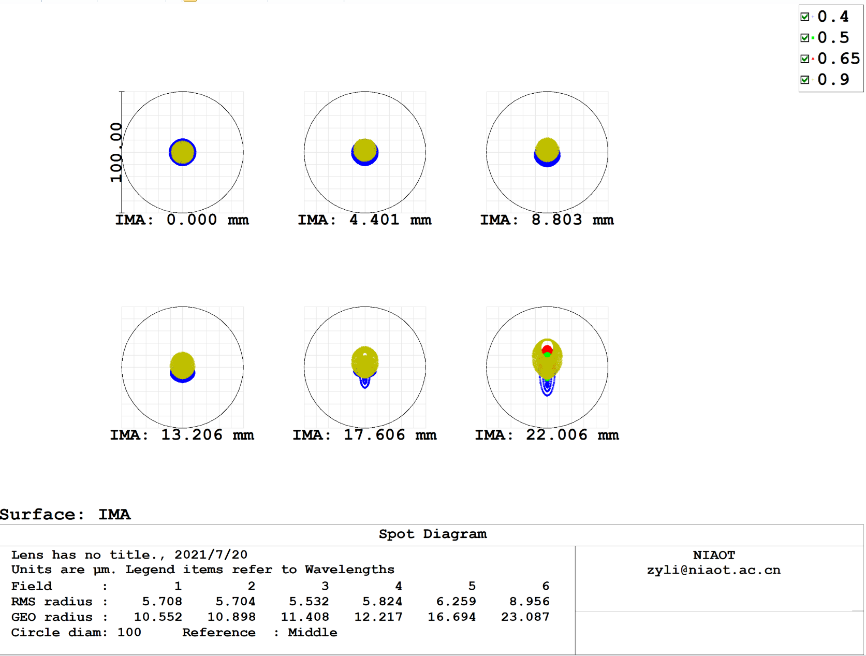}
   \caption{Tested spot diagram of the Mini-SiTian Telescope.}
   \label{Fig9:TestedSpotDiagram}
   \end{figure}

\begin{figure}
   \centering
   \includegraphics[width=12cm, angle=0]{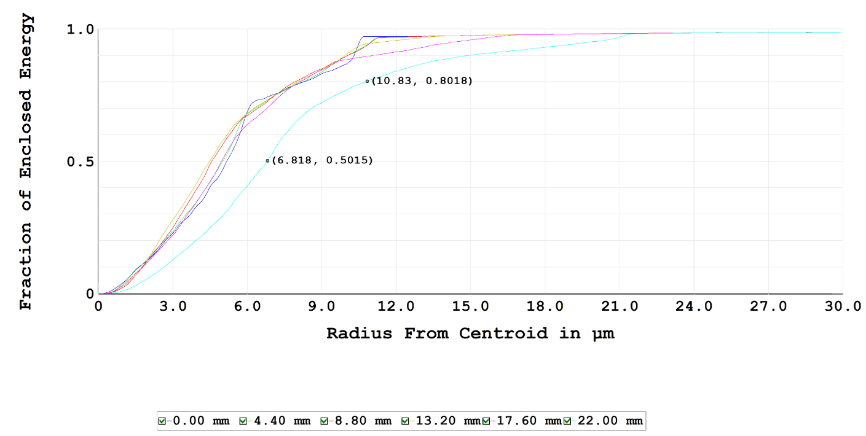}
   \caption{Tested diffraction encircled energy of the Mini-SiTian Telescope.}
   \label{Fig10:TestedEE}
   \end{figure}

Image quality, pointing accuracy, and tracking ability were tested in Nanjing, China, in July 2021. With manufacturing error incorporated into the optical design, the EE80 diameter of the optical system is about 5.1 arcsecs, and the FWHM is less than 3.4 arcsecs. The tested spot diagram and tested diffraction encircled energy diagram are shown in Figures ~\ref{Fig9:TestedSpotDiagram} and ~\ref{Fig10:TestedEE}. 

\begin{figure}
   \centering
   \includegraphics[width=12cm, angle=0]{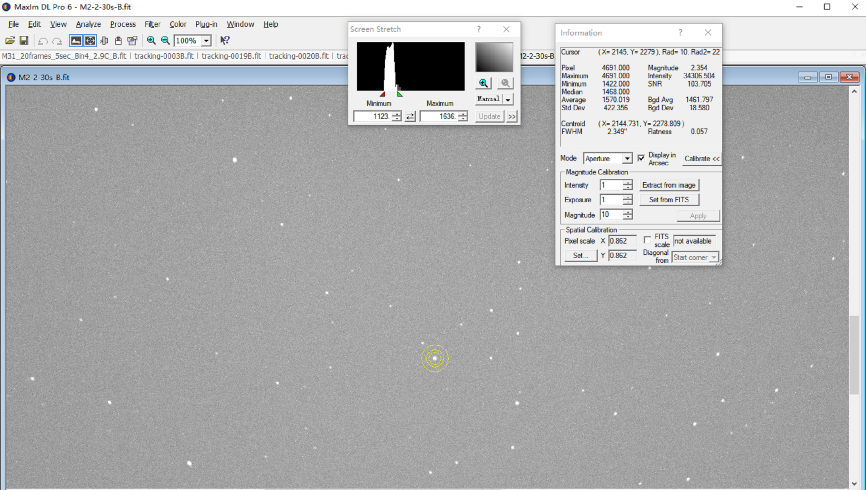}
   \caption{The tested minimum FWHM of the image is 2.35 arcsec.}
   \label{Fig11:TestedFWHM}
   \end{figure}

To evaluate the imaging performance, the telescope features a ZWO ASI6200MM Pro sCMOS camera mounted on its back. The size of the camera sensor is 36 mm $\times$ 24 mm, corresponding to FOV of 2.3$^{\circ}$ $\times$ 1.53$^{\circ}$. The tested minimum FWHM is 2.35 arcsecs, as Figure ~\ref{Fig11:TestedFWHM} shows.

\subsection{Pointing and Tracking Abilities}

\begin{figure}
   \centering
   \includegraphics[width=5cm, angle=0]{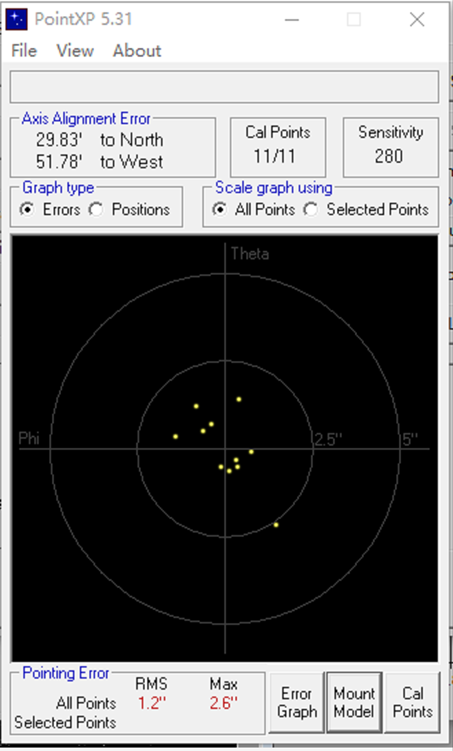}
   \caption{The tested RMS of pointing error is 1.2 arcsecs.}
   \label{Fig12：TestedRMS}
   \end{figure}

Pointing modeling is done before testing observations. The maximum pointing error is 2.6 arcsecs, and the RMS value is 1.2 arcsecs, as Figure ~\ref{Fig12：TestedRMS} shows. The pointing error is much lower than the requirement ($<$10 arcsecs). 

\begin{figure}
   \centering
   \includegraphics[width=12cm, angle=0]{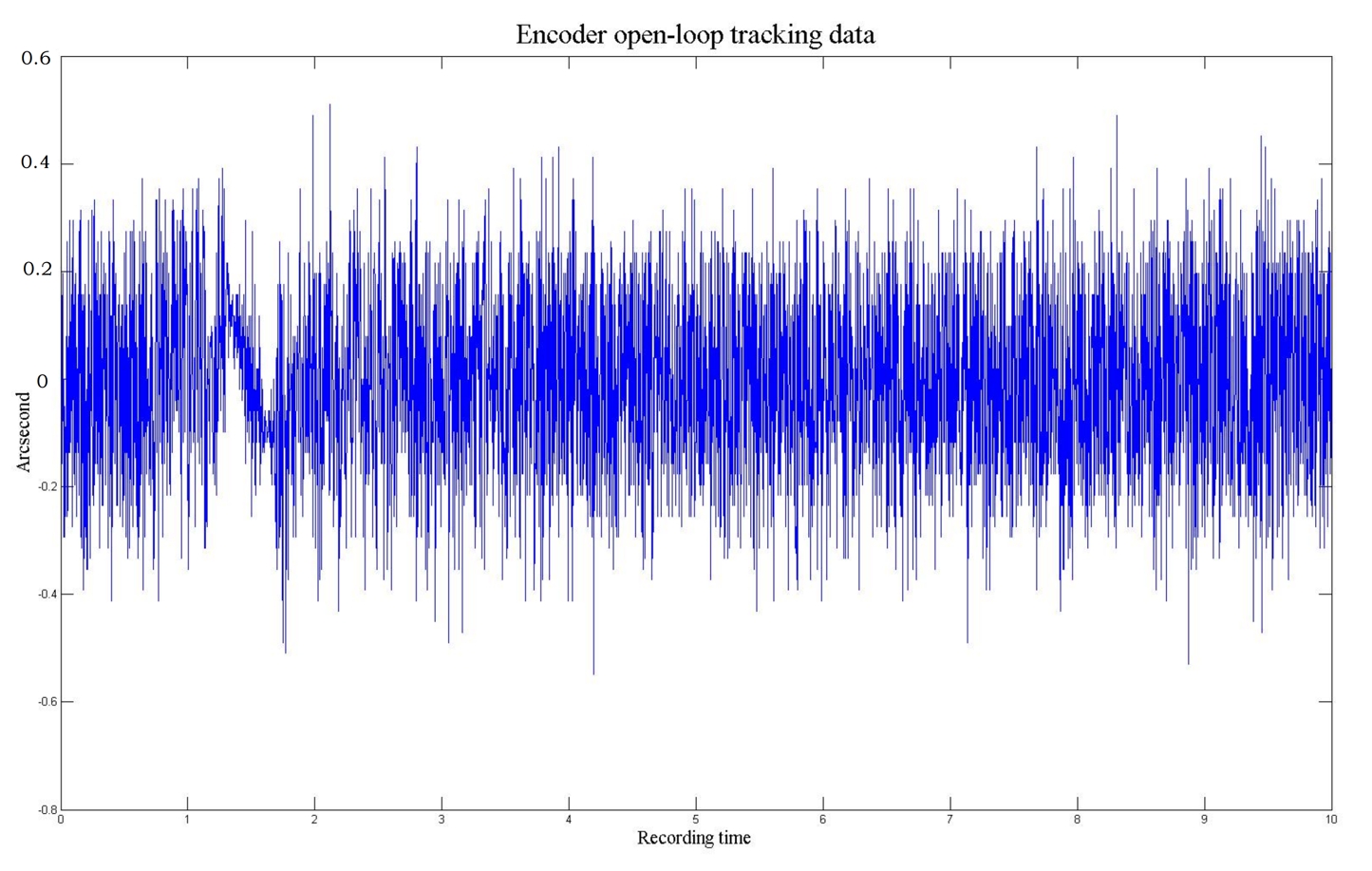}
   \caption{Tracking accuracy of the Mini-SiTian Telescope when observing a star.}
   \label{Fig13:TrackingAccuracy}
   \end{figure}

\begin{figure}
   \centering
   \includegraphics[width=12cm, angle=0]{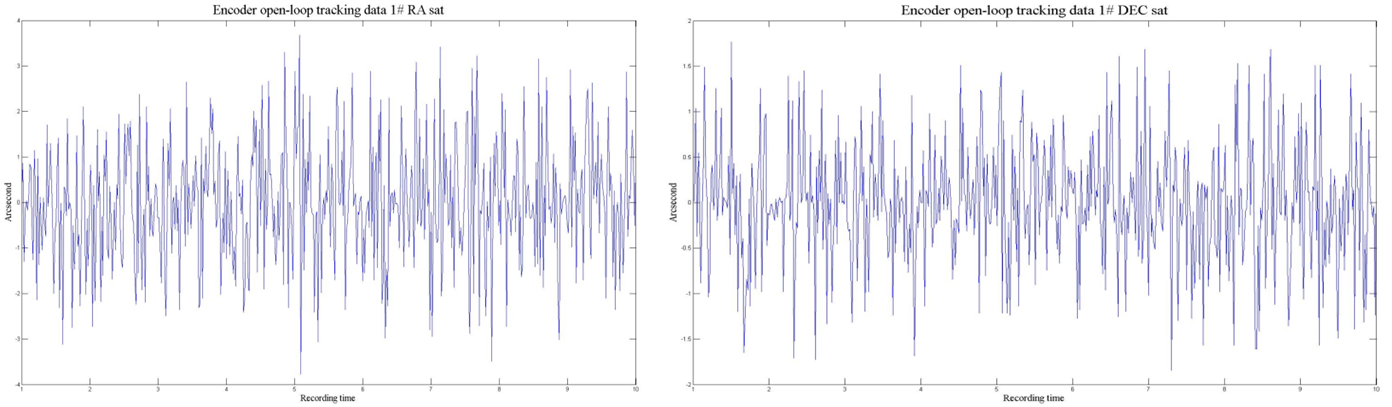}
   \caption{Tracking accuracy of the Mini-SiTian Telescope when observing a satellite.}
   \label{Fig14:Satellite}
   \end{figure}

To test the tracking ability, a star is observed by the telescope without any guide stars. Figure ~\ref{Fig13:TrackingAccuracy} shows the tracking ability when observing a star in 10 minutes. RA RMS is 0.175 arcsecs and DEC RMS is 0.03 arcsecs. The tracking accuracy of both axes is lower than the design requirements (0.5 arcsecs). A satellite is also observed to test the tracking accuracy. As Figure ~\ref{Fig14:Satellite} shows, RA RMS is 1.297 arcsecs, and DEC RMS is 0.660 arcsecs. The result shows the perfect tracking ability of the Mini-SiTian Telescope Array. 

\subsection{Commissioning}

\begin{figure}
   \centering
   \includegraphics[width=12cm, angle=0]{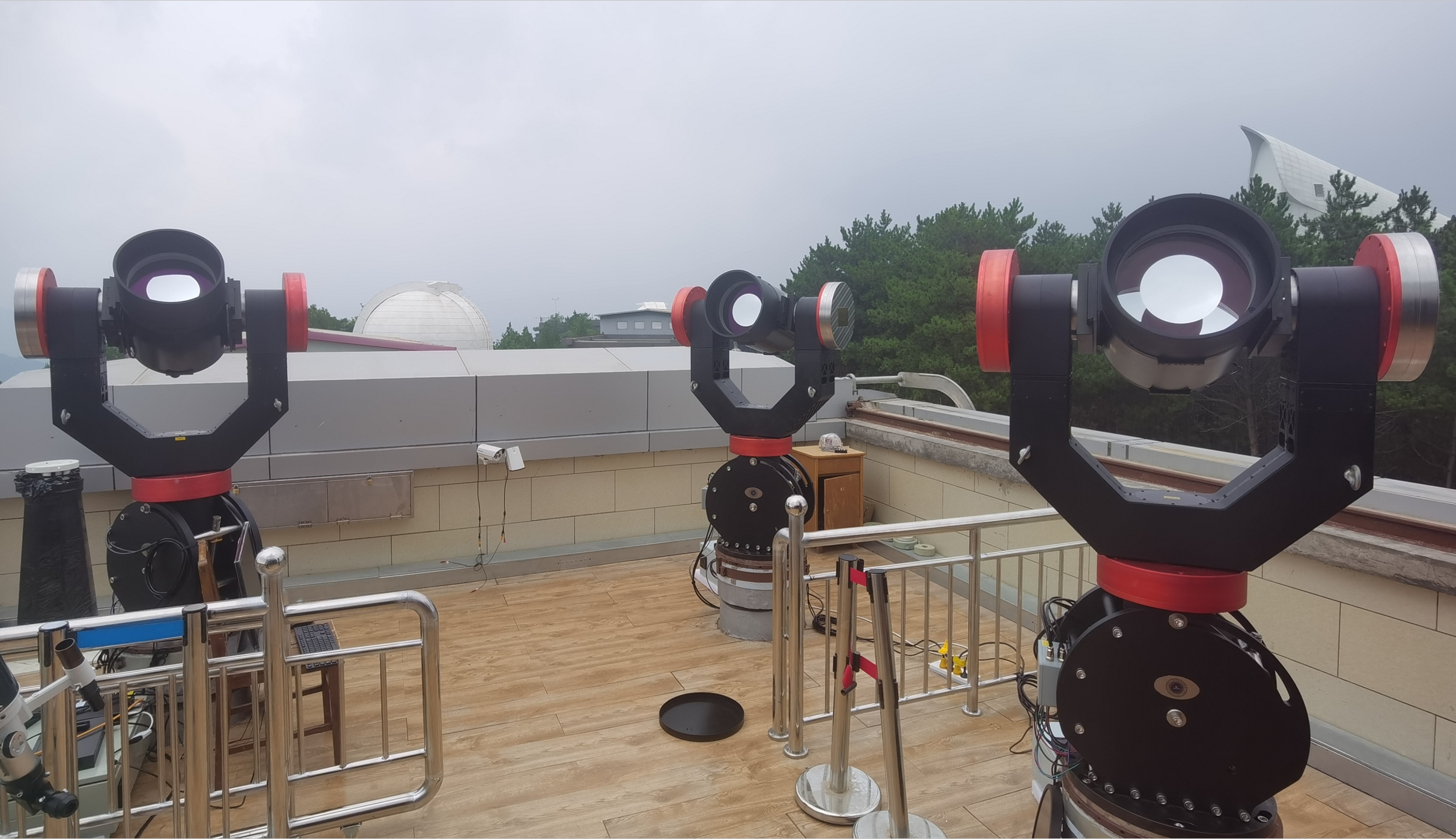}
   \caption{The Mini-SiTian Telescope Array commissioning at Xinglong Observatory.}
   \label{Fig15:MiniSiTianXinglong}
   \end{figure}

Mini-SiTian Telescope Array was transported to Xinglong Observatory of the National Astronomical Observatories, in August 2021. The Xinglong Observatory is located at $40^{\circ}23^{\prime}39^{\prime\prime}$N, $117^{\circ}34^{\prime}30^{\prime\prime}$E, with $\sim$900m average altitude. The mean and median seeing values of the Xinglong Observatory are $1.9^{\prime\prime}$ and $1.7^{\prime\prime}$, respectively. Most of the time, the sky brightness is about 21.1 mag/arcsec$^{2}$ in $V$ band at the zenith. The mounted Mini-SiTian Telescope Array at Xinglong Observatory is shown in Figure ~\ref{Fig15:MiniSiTianXinglong}. 

\section{Conclusions}

The Mini-SiTian Telescope Array is the demonstration platform for the SiTian project. Three telescopes are combined in one node to realize multi-color observations. The 300 mm F/3 telescope has an FOV of 5 degrees, which is specially designed for sky survey imaging. Commissioning in Xinglong Observatory, the software pipeline and control strategy have been successfully demonstrated and verified. The Mini-SiTian Telescope Array has proved to be a perfect platform for technical research and educational purposes.

\begin{acknowledgements}
The SiTian project is a next-generation, large-scale time-domain survey designed to build an array of over 60 optical telescopes, primarily located at observatory sites in China. This array will enable single-exposure observations of the entire northern hemisphere night sky with a cadence of only 30-minute, capturing true color (gri) time-series data down to about 21 mag. This project is proposed and led by the National Astronomical Observatories, Chinese Academy of Sciences (NAOC). As the pathfinder for the SiTian project, the Mini-SiTian project utilizes an array of three 30 cm telescopes to simulate a single node of the full SiTian array. The Mini-SiTian has begun its survey since November 2022. The SiTian and Mini-SiTian have been supported from the Strategic Pioneer Program of the Astronomy Large-Scale Scientific Facility, Chinese Academy of Sciences and the Science and Education Integration Funding of University of Chinese Academy of Sciences. This work is supported by the National Key Basic R\&D Program of China via 2023YFA1608304. 
\end{acknowledgements}

\bibliographystyle{raa}
\bibliography{ms2024-0399}

\label{lastpage}

\end{document}